\documentclass{rrparticle}
\usepackage{graphicx}

\newcommand{\miktex}{\hbox{Mik\kern-.15em\TeX}}

\newcommand{\ket}[1]{| #1 \rangle}

\newcommand{\etal}{{\textit{et al.}}}
\newcommand{\ignore}[1]{}

\title{Construction and characterization of a Sagnac-based entangled-photon source} 
\author[1]{Vasile-Lauren\c tiu Dosan}
\author[2,3]{Andrei Naz\^iru}
\author[4,5]{Mona Mih\u ailescu \footnote{C\lowercase{orresponding author}}}
\author[6]{Radu Ionicioiu}

\affil[1]{Faculty of Applied Sciences, University POLITEHNICA of Bucharest, 060042 Bucharest, Romania, email: {\em laurentiu.dosan@protonmail.com}}
\affil[2]{Extreme Light Infrastructure – Nuclear Physics, 077125 M\u{a}gurele-Bucharest, Romania, email: {\em andrei.naziru@eli-np.ro}}
\affil[3]{Physics Doctoral School, Bucharest University, 077125 M\u{a}gurele-Bucharest, Romania}
\affil[4]{Department of Physics, Faculty of Applied Sciences, University POLITEHNICA of Bucharest, 060042 Bucharest, Romania, email: {\em mona.mihailescu@physics.pub.ro}}
\affil[5]{Centre for Research in Fundamental Sciences Applied in Engineering, University POLITEHNICA of Bucharest, 060042 Bucharest, Romania}
\affil[6]{Horia Hulubei National Institute of Physics and Nuclear Engineering, 077125 Bucharest--M\u agurele, Romania, email: {\em r.ionicioiu@theory.nipne.ro}}


\begin{document}
\maketitle

\begin{abstract}

Entangled photons are crucial for the development of quantum technologies and especially important in quantum communications. Hence it is paramount to have a reliable, high-fidelity source of entangled photons. Here we describe the construction and characterization of a polarization-entangled photon source. We generate maximally-entangled Bell states using a type-II PPKTP nonlinear crystal inside a Sagnac interferometer. We characterize the source in terms of brightness, visibility, Bell-CHSH test and we perform quantum-state tomography of the density matrix. Our source violates Bell-CHSH inequality $|S|\le 2$ by $n=22$ standard deviations. With visibilities up to ${\cal V}= 98.9\%$ and fidelity ${\cal F}= 97\%$, our source is highly competitive, with state-of-the-art performance. To the best of our knowledge, this is the first source of entangled-photons designed and build in Romania and as such, represents an important step for the development of quantum technologies in our country. We envisage that our results will stimulate the progress of quantum technologies in Romania and will educate the future generation of quantum engineers.

\end{abstract}

\section{Introduction}

The new generation of quantum technologies use counter-intuitive aspects of quantum mechanics like superposition and entanglement \cite{2qrev, RI_rrp}. Entanglement is a key resource for quantum technologies and has several applications in quantum communications, quantum computing and quantum imaging. As such, a robust source of entangled photons is essential for any further development of the field.

Recently, entanglement generation techniques have been progressing considerably. A widely-used method to generate polarization-entangled photon pairs is spontaneous parametric down-conversion (SPDC) in nonlinear crystals \cite{Anwar_2021}. In bulk optics, one of the most stable sources of polarization-entangled photons use a Sagnac-interferometer setup. Sagnac-based quantum sources are widespread due to their phase-stability, versatility and the use of a single-cristal configuration \cite{Kim_2006, arxiv.2007.05095}. Nevertheless, this scheme requires polarization optics for different wavelengths (pump and down-converted photons) which makes the alignment process difficult. 

Due to its stability, an entanglement source based on the Sagnac configuration has been installed on the first quantum-communication satellite (Micius) \cite{Micius_rmp} and used to demonstrate satellite-to-ground quantum key distribution \cite{Micius_QKD}, entanglement distribution over 1200 km \cite{Micius_entang} and ground-to-satellite quantum teleportation \cite{Micius_teleport}.

In this article we describe the experimental realization of a Sagnac-based entanglement source and we characterize its output state. Our polarization-entangled photon source can be tuned to generate one of two maximally-entangled Bell states:
\begin{equation}
    \ket{\Psi^\pm}_{AB} = \frac{1}{\sqrt 2} (\ket{H_A V_B} \pm \ket{V_A H_B})
\end{equation}
where $H (V)$ labels horizontal (vertical) polarization and $A$ and $B$ the photons.

The structure of the article is the following. In Section 2 we discuss the architecture of the entanglement source and its main components. We briefly describe the alignment procedure, which is essential for obtaining a high-fidelity entangled state. In Section 3 we characterize the source and the output state in terms of several parameters (spectral brightness, visibility, state fidelity, purity, Bell parameter etc) and we perform the full quantum state tomography of the density matrix. Finally, we conclude in Section 4.

\section{Experimental setup}

\begin{figure}[t]
    \centering
    \includegraphics[width=12cm]{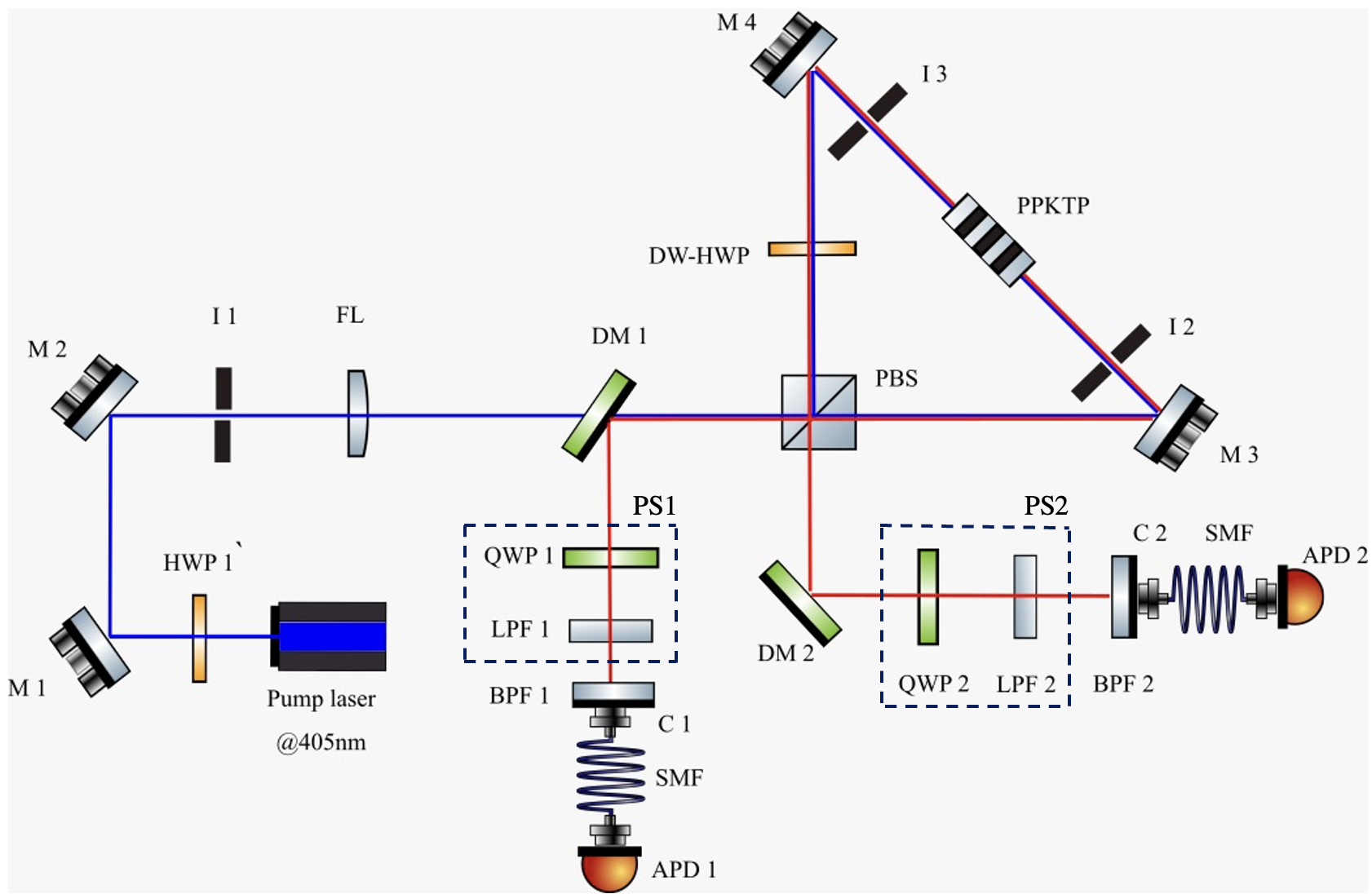}
    \caption{Sagnac-based source of polarization-entangled photons. The pump laser (down-converted photons) are in blue (red). HWP: half-wave plate; DW-HWP: dual-wavelength HWP; QWP: quarter-wave plate; M: mirror; I: iris; FL: focusing lens; DM: dichroic mirror; PBS: polarization beam splitter; LPF: linear polarizer filter; BPF: bandpass filter; C: fiber coupler; PS1,2: polarization stages (QWP, LPF); SMF: single-mode fiber;  APD: avalanche photo diode.}
    \label{fig:experimental setup Sagnac}
\end{figure}

In this section we describe in detail the configuration of the entangled-photons source. Briefly, it consists of a laser (405 nm) who pumps bidirectionally a nonlinear PPKTP crystal situated in a Sagnac interferometer, Fig.~\ref{fig:experimental setup Sagnac}. The source generates polarization-entangled photon pairs by SPDC from 405 nm to 810 nm.

Specifically, the source is based on collinear type-II quasi phase matching (QPM) on a 15\,mm-long PPKTP crystal (Raicol) with a grating period of $10 \, \mu$m. The source is pumped with a free-space continuous-wave (CW) laser diode at $\lambda_p =$ 405\,nm (Ondax compact laser module). The photons emerging from the laser diode are vertically-polarized. A half-wave plate (HWP) is set at $22.5^{\circ}$ to prepare the superposition $\ket{+}= \frac{1}{\sqrt 2} (\ket{H} + \ket{V})$.

The pump beam is focused in the middle of the crystal by a convex lens (FL, $f= 75$ cm). The beam passes a dichroic mirror (DM1) which transmits 405\,nm and reflects 810\,nm. The $H$ component is transmitted by the polarization beam splitter (PBS) and enters the Sagnac loop in counter-clockwise direction, while the $V$ component is reflected by the PBS and follows a clockwise path. 

Inside the loop (in the clockwise arm) we place a dual-wavelength half-wave plate (DW-HWP) for 405\,nm and 810\,nm, which is set to $45^{\circ}$, such that $H$- and $V$-polarizations are swapped, $H\rightarrow V$, $V\rightarrow H$. This ensures that the crystal is pumped (at 405\,nm) from both directions by $H$-polarized photons. This is crucial, since SPDC occurs in the crystal only for $H$-polarized pump photons. At the exit, both $H$ and $V$ pump photons (405\,nm) will exit the loop and will be transmitted back by the dichroic mirror (DM). The PPKTP crystal is kept inside a Peltier oven at a controlled temperature (between 20$^\circ$ and 60$^\circ$C) required for QPM. Inside the crystal SPDC occurs and a part of $H$-polarized pump photons are down-converted into pairs of 810\,nm with orthogonal polarization. 

The 810 nm photons travel back to the PBS. After the PBS, the down-converted photons exit the loop via the two outputs. The first output consists of a $HV$ pair that are reflected by a dichroic mirror (DM2) to the polarization stage (PS2). The second output emits $VH$ pairs which are also reflected by another dichroic mirror DM1 to the polarization stage PS1. In other words, in the Sagnac interferometer the two paths are coherently superposed and, if the pump is in the $\ket{+}$ state, both paths will have the same amplitude. 

After the two polarization stages PS1/PS2, the photons are coupled into single-mode fibers (SMFs) using opto-couplers (OCs). A 810 $\pm$ 10\,nm bandpass filter (BPF) is placed before each OC. The BPF are used to minimize the noise and to filter better the 405 nm photons. The FC/PC SMFs are connected to a quED quTools module which contains two Si-based avalanche photo diodes (Si-APDs, quantum efficiency $\eta= 30\%$) and a coincidence module; the coincidence window width is set to 30\,ns.

The main operating parameters of the source are shown in Table \ref{tab:paramsheet}.

\begin{table}[h!]
\centering
\begin{tabular}{lccc}

\textbf{Parameters} & \textbf{min} & \textbf{typ} & \textbf{max} \\
\hline
 
Laser diode input current & 50\,mA & 80\,mA & 95\,mA \\
\hline
 
Laser diode stabilization time & 0.5 hrs& 2 hrs& -- \\
\hline

Pump wavelength & 404\,nm & 405\,nm & 406\,nm \\
\hline

Output (down-converted) wavelength & -- & 810\,nm & -- \\
\hline

Oven temperature & 20\,$^\circ$C & 33.2\,$^\circ$C & 60\,$^\circ$C \\
\hline

Single count rate & 1\,kHz & 30\,kHz& 50\,kHz \\
\hline

Dark count rate & -- & 500\,Hz & -- \\
\hline

Detection efficiency & -- & 30\,\% & -- \\
\hline

Coincidence window length & -- & 30\,ns & -- \\
\hline

\end{tabular}
\caption{Operating parameters of the quantum source, with min, max and typical values.}
\label{tab:paramsheet}
\end{table}

Building and aligning the source is non-trivial and several steps are required in order to obtain a highly-entangled output state. We now briefly describe the alignment procedure.

The construction of the setup took several months because the alignment process was difficult. From our experience, a few key features of the alignment procedure are to: (i) visualize fringes at 405 nm before inserting the PPKTP crystal inside the loop; (ii) ensure the proper degrees of freedom for each optical component; (iii) keep the crystal at the right temperature in order to generate the target quantum state; (iv) position precisely the PPKTP crystal in the middle of the interferometer.

The setup misaligns easily between measurements performed in different days. In this case, a few hints are: the distances between the optical components should be as small as possible; a constant temperature and humidity should be maintained in the lab; the setup should be on a vibration-free optical table; the lab windows should be covered to minimize the dark counts. In order to re-align the setup, we have to do the following steps: improve the coupling; remove the LPFs and achieve a constant coincidence rate for each HWP1 position using the PBS and M2,3,4; find a proper crystal temperature for good entanglement visibilities.

\section{Source characterization}

The next important step after aligning the quantum source is to characterize its output in terms of brightness, entanglement, state fidelity, purity etc. These parameters will be described in the present section. To begin with, the quality of the alignment is directly reflected in the characteristics of the output quantum state.

From the coincidence-to-single-photon coupling ratio ($\eta_{C} = \frac{S_C}{\sqrt{S_i S_s}}\approx 9.87\,\%$, where $S_C$, $S_i$, $S_s$ are the coincidence, idler and signal rates, respectively), the detection efficiency ($\eta_D \approx 30\,\%$) and the single-mode bandwidth of the down-converted photons ($\Delta \lambda = \frac{5.52 \cdot 10^{-3}}{L} = 0.4\,\mathrm{nm}$, where $ L = 15\,\mathrm{mm}$ is the crystal length), we estimated the spectral brightness, $S \approx 62 500 \,(\rm mW \cdot s \cdot nm)^{-1}$ \cite{tenger_2005, Fedrizzi:07}. This gives the number of entangled photon pairs per second, per mW of pump power and per nm of wavelength bandwidth. By integrating over the wavelength bandwidth, we obtain the source brightness $B= 25000\,(\rm mW \cdot s)^{-1}$, i.e., the number of entangled photon pairs per second, per mW of pump power.

\begin{figure}[t]
    \centering
    \includegraphics[width=13cm]{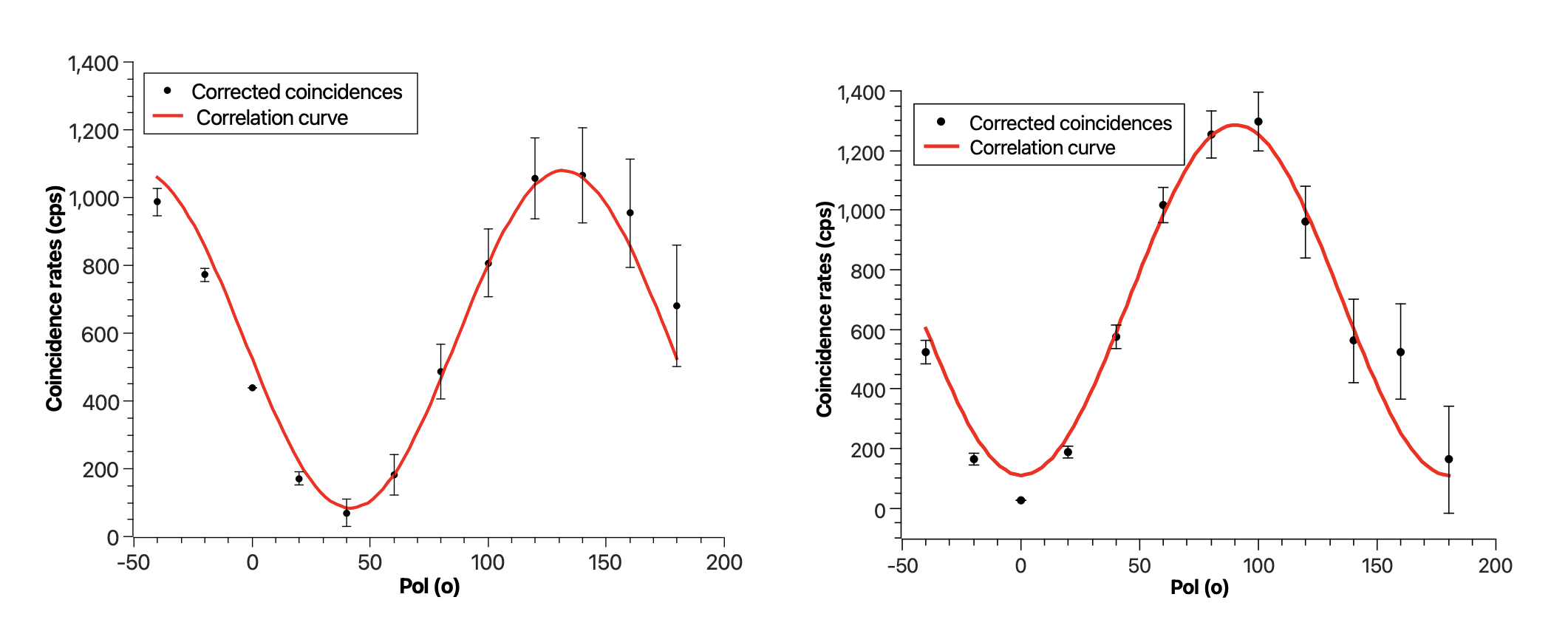}
    \caption{Correlations curves for $+/-$ basis (left) and $H/V$ basis (right).}
    \label{fig:correlation curves}
\end{figure}

One of the most important parameters is the entanglement of the output state. The easiest way to test the entanglement is to measure the correlation curves in two mutually-unbiased bases. This can be done by keeping a linear polarizer filter at a fixed angle ($\theta_1$) while changing the orientation ($\theta_2$) of the other one. For the $H/V$ basis $\theta_1 = 0^{\circ}$ and for $+/-$ basis $\theta_1 = 45^{\circ}$. The entanglement visibility is given by:
\begin{equation}
    {\cal V} = \frac{C_{\rm max} - C_{\rm min}}{C_{\rm max} + C_{\rm min}}
\end{equation}
where $C_{\rm min/max}$ is the minimum/maximum coincidence rate determined from the correlation curves (Fig.~\ref{fig:correlation curves}). From our experimental data, the visibilities in the two bases are ${\cal V}_{H/V} = (98.9 \pm 0.4) \,\%$ and ${\cal V}_{+/-} = (93.7 \pm 1.1) \,\%$.

Another way to quantify the entanglement is to test the violation of Bell-CHSH inequality. As shown by Bell and others, any local hidden-variable theory satisfies \cite{PhysicsPhysiqueFizika.1.195, PhysRevLett.23.880, PhysRevA.49.3209}:
\begin{equation}
|S| \leq 2
\label{Bell_ineq}
\end{equation}
where the Bell-CHSH parameter $S$ is defined as
\begin{equation}
S = E(\alpha, \beta) - E(\alpha, \beta') + E(\alpha', \beta) + E(\alpha', \beta')
\end{equation}
Here $\alpha, \alpha', \beta, \beta'$ are different polarizer angles and $E(\alpha, \beta)$ is the expectation value of joined polarization measurements (of the two photons) along directions $\alpha$ and $\beta$:
\begin{equation}
E(\alpha, \beta) = \frac{C(\alpha, \beta) + C(\alpha_{\perp}, \beta_{\perp})- C(\alpha, \beta_{\perp})- C(\alpha_{\perp}, \beta)}{C(\alpha, \beta) + C(\alpha_{\perp}, \beta_{\perp})+C(\alpha, \beta_{\perp})+C(\alpha_{\perp}, \beta)}
\end{equation}
where $C(\alpha,\beta)$ is the coincidence rate for polarizer angles $\alpha$ and $\beta$; $\alpha_{\perp} = \alpha + 90^{\circ}$ and $\beta_{\perp} = \beta + 90^{\circ}$.

However, in quantum mechanics the inequality \eqref{Bell_ineq} is violated for certain entangled states. The maximum violation is obtained for maximally-entangled (Bell) states and for polarizer angles $(\alpha, \beta, \alpha', \beta')= (0, 22.5^\circ, 45^\circ, 67.5^\circ)$. In this case, quantum mechanics gives
\begin{equation}
|S|_{QM} = 2 \sqrt{2}\approx 2.828
\end{equation}
Thus, to test Bell-CHSH inequality we need to measure polarization correlations for joint measurements of the two-photons state at different polarizer angles $\alpha, \beta$. 

We have measured the 16 coincidence rates for the previous angles $(\alpha, \beta, \alpha', \beta')= (0, 22.5^\circ, 45^\circ, 67.5^\circ)$ for our source. We have obtained 
\begin{equation}
|S|= 2.684 \pm 0.03
\end{equation}
Thus our source violates the Bell-CHSH inequality by more than $n_{\Delta}= 22$ standard deviations, with $n_{\Delta} = (\overline{|S|}-2)/ \Delta S$, $\overline{|S|} = 2.684, \Delta S = 0.03$. Consequently, there is reliable and strong evidence that our source generates highly-entangled states violating Bell-CHSH inequality \eqref{Bell_ineq}.

\begin{figure}[t]
    \centering    \includegraphics[width=12cm]{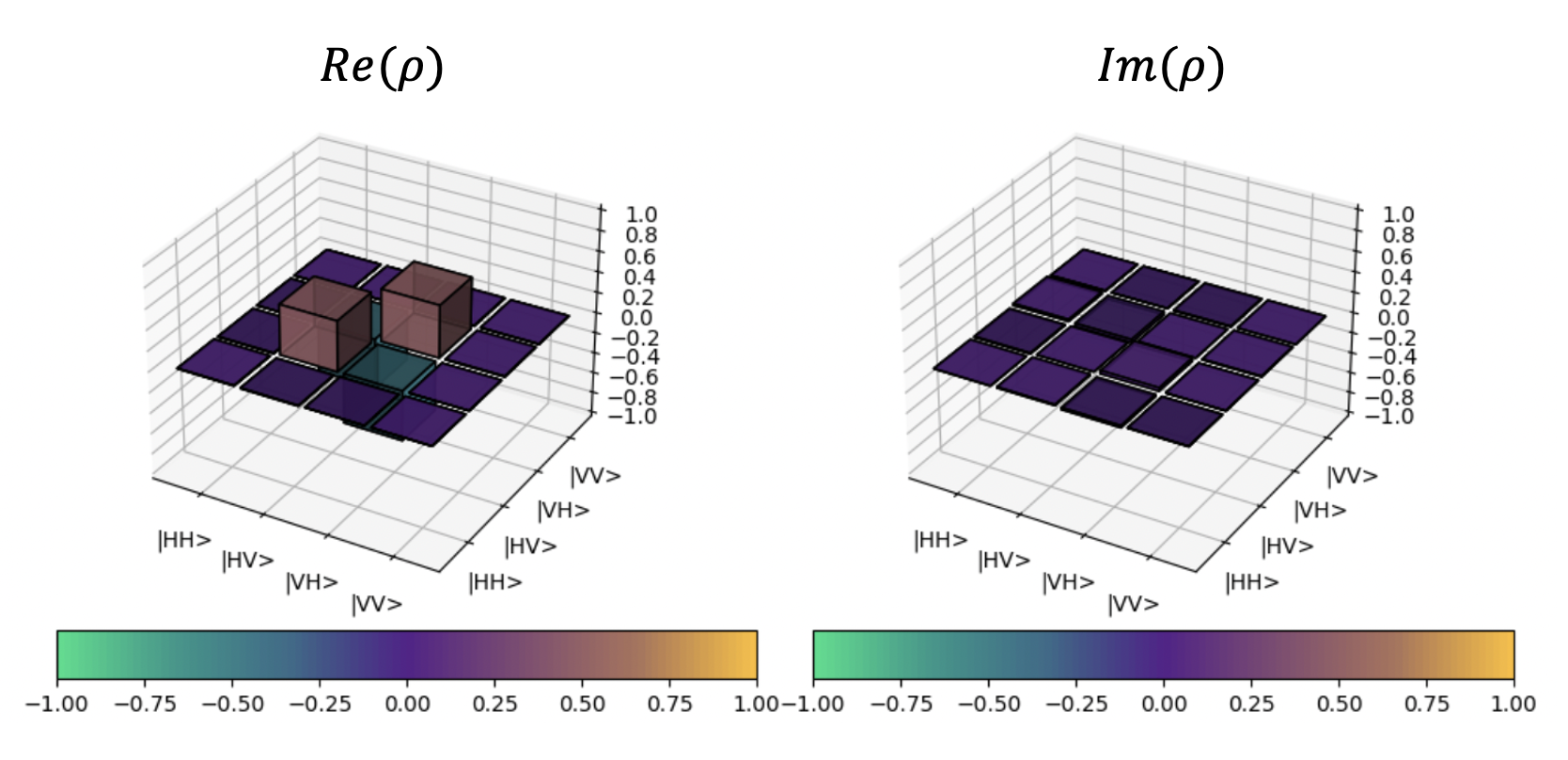}
    \caption{The reconstructed (maximum-likelihood) density matrix, with the real part (left) and imaginary part (right).}
    \label{fig:density_matrix}
\end{figure}

To fully characterize the output state we performed quantum state tomography. This procedure reconstructs the full density matrix of a state from the set of coincidence measurements for different polarization measurements \cite{PhysRevA.64.052312}. This procedure uses: (i) coincidence counts for measurements in different bases ($HH, HV$ etc), and (ii) a numerical optimization technique known as maximum-likelihood (ML) \cite{Motazedifard_2021}.

Using our experimental data for coincidence counts in different bases, we have reconstructed the maximum-likelihood density operator $\rho_{ML}$ using an open-source (online) tool from Kwiat group (Illinois University) \cite{interface}. Our reconstructed maximum-likelihood density operator $\rho_{ML}$ is given below, see Fig.~\ref{fig:density_matrix}.

\begin{equation}
\small
\rho_{ML} = 10^{-3} \cdot
\begin{bmatrix}
4.24 & -6.17 - 9.42 \rm i & -0.275 + 19.6\rm i & 1.26 + 3.78 \rm i \\
-6.17 + 9.42 \rm i & 486 & -487 - 032 \rm i & 0.754 - 0.574 \rm i \\
-0.275 + 19.6 \rm i & -487 - 32 \rm i & 505 & 7.02 - 7.84 \rm i \\
1.26 + 3.78 \rm i & 0.754 + 0.574 \rm i & 7.02 - 7.84 \rm i & 4.02
\end{bmatrix}
\end{equation}

For a mixed state the amount of entanglement is quantified by the concurrence $C$ \cite{concurrence}; $C=0$ for separable states and $C=1$ for maximally-entangled states. For our source we have:
\begin{equation}
C= 0.951
\end{equation}

The source is aligned to generate the $\ket{\Psi^{-}}$ state. The fidelity $\cal F$ measures how close is the output state from the target state:
\begin{equation}
     {\cal F} (\rho_{ML}, \rho_{\ket{\Psi^-}}) = \mathrm{tr}^2 \left. \sqrt{ \rho_{ML}^{1/2}\ \rho_{\ket{\Psi^-} }\ \rho_{ML}^{1/2} }\right. = 0.9743
\end{equation}
This shows that our output state is very close to the desired maximally-entangled state $\ket{\Psi^-}$. Another relevant parameter is the purity:
\begin{equation}
    P= {\rm tr} (\rho_{ML}^2) = 0.953
\end{equation}
The purity varies between $\frac{1}{4}$ (for the totally mixed state) to 1 (for pure states). 

The main output parameters of our source are summarised in Table \ref{tab:state_characteristics}.

\begin{table}[t]
\centering
\begin{tabular}{lr}
\textbf{Parameter} & \textbf{value} \\
\hline
 
Brightness & 2.5$\times 10^4$\,$\rm (mW \cdot s)^{-1}$\\
\hline

Spectral brightness & 6.25$\times 10^4$\,$\rm (mW \cdot s \cdot nm)^{-1}$ \\
\hline

${\cal V}_{H/V}$ visibility & $98.9 \pm 0.4 \, \%$ \\
\hline

${\cal V}_{+/-}$ visibility & $93.7 \pm 1.1 \, \%$ \\
\hline

Bell-CHSH parameter & $|S| = 2.684 \pm 0.03$ \\
\hline

Concurrence & $C= 0.951$ \\
\hline

Fidelity & ${\cal F}= 0.9743$ \\
\hline

Purity & $P= 0.953$ \\
\hline

\end{tabular}
\caption{The output parameters of our entangled-photons source.}
\label{tab:state_characteristics}
\end{table}

\section{Discussion and conclusions}

\begin{table}[h!]
\centering
\begin{tabular}{lccc}
 \textbf{Year} & \textbf{ref.}  &  \textbf{visibility ($\%$)} &\textbf{fidelity}\\
 \hline
 
{\bf 2022} & {\bf this work}  & {\bf 98.9} & {\bf 0.97}\\

2022 & \cite{Cai:22} & 98.2 & 0.96\\

2021 & \cite{Motazedifard_2021} & 99.9 & 0.97\\

2021 & \cite{Meraner_2021} & - & 0.99\\

2018 & \cite{Meyer-Scott:18} & 96.0 & 0.95\\

2018 & \cite{Terashima_2018} & 98 & 0.87\\
\hline

\end{tabular}
\caption{Performance of our source compared to state-of-the-art.}
\label{tab:state_of_the_art}
\end{table}

In this article we have described the experimental realization of a polarization-entangled photon source. Maximally-entangled states are generated through SPDC in a type-II nonlinear crystal inside a Sagnac interferometer. The quality of the alignment is directly reflected by the parameters of the output state. We have characterized the quantum source in terms of several parameters: brightness, visibility, Bell-CHSH test, concurrence and we have performed quantum state tomography of the density matrix.

A comparison between our source and state-of-the-art is given in Table \ref{tab:state_of_the_art}. As we can see, our source is highly competitive, and in some cases equals or exceeds existing experiments, in terms of visibility and fidelity.

To the best of our knowledge, this is the first source of entangled-photons designed and build in Romania. As such, it represents an important step for the development of quantum technologies in our country. In the future, our source can be used to demonstrate quantum teleportation and entanglement-based quantum key distribution. We envisage that our results will stimulate the progress of quantum technologies in Romania and will educate the future generation of quantum engineers.

\noindent {\bf Acknowledgements.} The authors acknowledge support from a grant of the Romanian Ministry of Research and Innovation, PCCDI-UEFISCDI, project number PN-III-P1-1.2-PCCDI-2017-0338/79PCCDI/2018 (QUTECH-RO), within PNCDI III. R.I. acknowledges support from PN 19060101/2019-2022.


\end{document}